\documentclass[letterpaper, 12 pt]{article}
\usepackage[a4paper, total={6in, 8in}]{geometry}
\usepackage{graphicx} 
\usepackage{amsmath}
\usepackage{graphicx}
\usepackage{cancel}
\usepackage{lipsum}
\usepackage{cite}
\usepackage{amssymb}
\usepackage{xcolor}

\title{A note on analogue semi-classical gravity in (1+1) dimensions}
\author{ Akshat Pandey\footnote{apandey.physics@gmail.com} \\ \normalsize Department of Physics, Shiv Nadar Institution of Eminence \\ \normalsize Greater Noida, Uttar Pradesh-201314, India.}

\date{}
\begin{document}

\maketitle
\begin{abstract}
    Acoustic spacetimes have been known to offer analogue models for black hole physics and cosmology. Within this context, aspects of analogue quantum field theories in curved spacetime are studied. In particular some new comments have been made on the analogue Hawking temperature including a quick derivation of the result. Further, analogue cosmology is explored, within which, an acoustic version of the Parker-Toms model is proposed and the corresponding quantities have been calculated. The limits of the acoustic analogue are emphasised.
\end{abstract}

\section{Introduction}
In 1981, Unruh \cite{unruh}, in his paper on experimental black hole evaporation, predicted a thermal spectrum of sound waves from the sonic horizon of trans-sonic fluid flows. Further, in 1997, Visser formalised the theory of acoustic black holes\cite{visser}. He showed as a theorem that if there is a non-relativistic fluid governed by the Euler equation and the continuity equation, which in addition is inviscid, irrotational and barotropic. Then sound propogation in such a fluid is described by a massless minimally-coupled scalar field, propagating in a (3+1) Lorentzian acoustic metric:
\begin{equation}
    g_{\mu \nu} \equiv \displaystyle\frac{\rho_{0}}{c}\left[\begin{array}{ccc}
-\left(c^{2}-v_{0}^{2}\right) & \vdots & -v_{0}^{j} \\
\ldots \ldots \ldots \ldots & \cdot & \ldots \ldots \\
-v_{0}^{i} & \vdots & \delta_{i j}
\end{array}\right] 
\end{equation}
Such that the equation for sound wave propagation is:
\begin{equation}
    \partial_{\mu}\left(\sqrt{-g} g^{\mu \nu} \partial_{\nu} \phi\right)=0
\end{equation}

Ever since, there has been a continual interest in studying acoustic black holes as analogue models for gravity, for example \cite{qnm, anacleto}. In particular these acoustic spacetimes offer analgoues for the kinematic effects of GR, that is, GR takes place in Lorentzian spacetimes. This includes, importantly, analogues to the effects predicted by applying quantum field theories (QFTs) to curved spacetimes. In fact, the physics of low-momentum phonons is completely equivalent to quantum field theories in curved spacetimes \cite{visser2}. 

In this article, we explore some of the features of the acoustic analogue, particularly in this context of applying QFTs to curved spacetimes. Since the acoustic metric is obtained from the equations of fluid mechanics, it is straightforward to obtain an effective (1+1) dimensional spacetime, by choosing the fluid velocity field to vanish in two of the three spatial dimensions. This is unlike GR where the (3+1) metric structure is obtained from the highly nonlinear Einstein equations in 4 spacetime dimensions and finding exact solutions that are effectively 2 dimensional is difficult to conceive. This implies, for example, that the acoustic analogue of a (3+1) dimensional black hole can be effectively (1+1) dimensional, as long as the fluid flow is kept trans-sonic. This is further advantageous as most of the calculations greatly simplify in (1+1) dimensions. 
Equipped with only a (1+1) dimensional acoustic metric, and a scalar field minimally coupled to it, we work out certain consequences analogous to results known within semi-classical gravity. For other works on analogue gravity in lower dimensions, see \cite{rinaldi, cadoni}

This article is organised as follows. In section 2, we sketch the construction of an effective (1+1) dimensional acoustic metric. In section 3, we revisit some aspects of analogue black hole physics and Hawking temperature, based on Visser's derivation of the same \cite{visserhawking}. In section 4, we explore certain consequences of a scalar field in an analogue FLRW background, in particular, the Parker-Toms model. We end with a summary in section 5.

\section{An effective 1+1 dimensional background}
The acoustic metric is:
\begin{equation}
    \mathrm{d} s^{2} =\frac{\rho_{0}(\vec{x},t)}{c_s(\vec{x},t)}\left[-c_s^{2} \mathrm{~d} t^{2}+\delta_{i j}\left(\mathrm{d} x^{i}-v_{0}^{i} \mathrm{~d} t\right)\left(\mathrm{~d} x^{j}-v_{0}^{j} \mathrm{~d} t\right)\right].
\end{equation}

This metric describes a Lorentzian acoustic spacetime to which sound, described by a scalar field $\phi$ is coupled to. Here both the fluid velocity $v_0(\vec{x},t)$ and sound velocity $c_s(\vec{x},t)$ are 3-vector fields.

The choice the velocity field fixes the metric and determines the background geometry. In order to have an effectively (1+1) dimensional spacetime, the velocity field is chosen in cartesian coordinates to be
\begin{equation}
    \vec{v_0}(z,t)=(0,0,v_z(z, t)).
\end{equation}
This choice of the velocity field ensures that the fluid flow is only in the $z$ direction. The $x$ and $y$ dependence is trivial and can be safely neglected to get an effective (1+1) dimensional metric.

The acoustic metric (1) thus becomes
\begin{equation}
     \mathrm{d} s^{2} =\frac{\rho_{0}(z,t)}{c_s(z,t)}\left[- (c_s^{2}-v_z^2)\mathrm{~d} t^{2}-2v_z\mathrm{d} z \mathrm{d}t+\mathrm{d} z^2 \right].
\end{equation}

There is still some freedom left in defining $\rho_{0}(z,t)$ and $c_s(z,t)$. Soon we will see that this will come in handy while defining and distinguishing analogues to black hole and cosmological spacetimes.

\section{Analogue Black hole physics}

In order to get a black hole like spacetime from (5), two further conditions are imposed.
\begin{enumerate}
    \item The speed of sound $c_s$ is chosen to be a constant. In this section, we work in units such that $c_s=1$.
    \item The background fluid velocity $v_z(z,t)$ is chosen to be explicitly time independent. This makes sense because, like the Schwarszchild solution, the acoustic black hole solution is stationary. Further, $v_z$  is required to be a monotonically decreasing in $z$ (within certain bounds of $z$ atleast), to ensure a trans-sonic flow. Notice that in the case of trans-sonic flow, $v_z$ would describe an acoustic black hole with horizon $z_H$ described as the surface where the speed of the fluid equals the speed of sound i.e. $v_{z_{H}}=1$
\end{enumerate}

Imposing these conditions on (5), 

\begin{equation}
    \mathrm{d} s^{2} =\rho_{0}\left[- (1-v_z^2)\mathrm{~d} t^{2}-2v_z\mathrm{d} z \mathrm{d}t+\mathrm{d} z^2 \right].
\end{equation}

At this point we notice that the line element (6) has a similar form to the Schwarzschild line element in Painlev\'{e} coordinates. We perform a coordiante transformation from the laboratory time($t$) to a Schwarszchild like time coordinate($\tau$) such that
\begin{equation}
    \mathrm{d} \tau = \mathrm{d} t + \frac{v_z}{1-v_z^2}\mathrm{d}z.
\end{equation}
This converts the line element (6) to a more convenient representation
\begin{equation}
    \mathrm{d} s^{2} =\rho_{0}\left[- (1-v_z^2)\mathrm{~d} \tau^{2}+ \frac{1}{(1-v_z^2)}\mathrm{d}z^2 \right].
\end{equation}
This can be called the transform canonical (1+1) dimensional acoustic black hole metric. Note, we work out with the above acoustic black hole solution obtained within non-relativistic fluid mechanics. For acoustic black holes obtained within relativistic fluids, see \cite{sin, neven, ana1, ana3}. The source of the curvature here is the inhomgenous flow of the fluid i.e. the background velocity field $\vec{v_z}$. As a check, we can see that the acoustic metric reduces to a Minkowski-like metric (upto a conformal factor) when the background fluid velocity becomes zero.

The position dependent conformal factor $\rho_0$ does not affect the Hawking temperature \cite{jacobson}, therefore we will simply neglect it from here on.

The line element can be written as 
\begin{equation}
    \mathrm{d} s^{2} =- (1-v_z^2)\mathrm{~d} \tau^{2}+ \frac{1}{(1-v_z^2)}\mathrm{d}z^2. 
\end{equation}

\subsection{The scalar field}
Sound propagation in acoustic spacetimes is described by the minimally coupled massless scalar field equation. Note, this is valid under the assumption that sound in the fluid is described as a linearised fluctuation of its dynamical quantities. Upon quantisation the scalar field gives rise to phonons.

The equation for sound propagation is
\begin{equation}
    \partial_{\mu}\left(\sqrt{-g} g^{\mu \nu} \partial_{\nu} \phi\right)=0.
\end{equation}
For the (1+1) dimensional metric, $\mu=(0,1)$. Expanding equation (9) this metric, we get
\begin{equation}
    -(\frac{1}{1-v_z^2})\partial_\tau ^2 \phi + \partial_z (1-v_z^2) \partial_z \phi + (1-v_z^2)\partial_z ^2 \phi =0.
\end{equation}
We will look at solutions of the form
\begin{equation}
    \phi (z, \tau)= \psi(z)e^{-i \omega \tau}.
\end{equation}
We also note that, for analogue static spacetimes, the surface gravity $K$ is defined as 
\begin{equation}
    K=\frac{dv_z}{dz}
\end{equation}
It is indeed the surface gravity of (analogue) black holes which, upto a constant, is identified with the (analogue) Hawking temperature \cite{bardeen}.

For the rest of this subsection we will solve for $\psi(z)$ in order to obtain its dependence on $K$ through a Boltzmann-like factor. This, in turn, will allow us to define a corresponding temperature.
Plugging (12) into (11), we get
\begin{equation}
    \frac{\omega ^2}{(1-v_z^2)}\psi + \partial_z (1-v_z^2) \partial_z \psi + (1-v_z^2)\partial_z ^2 \psi =0.
\end{equation}

In order to simplify this equation, we will make use of the tortoise coordinate $z_*$ such that
\begin{equation}
    \frac{dz_*}{dz}= \sqrt{-\frac{g_{zz}}{g_{00}}}=\frac{1}{1-v_z^2}.
\end{equation}
We further note that tortoise co-ordinates, by definiton, only cover the region outside the horizon \cite{carroll, waldgr}, i.e. ($v_z < 1$). Qualitatively, we see
\begin{equation}
    \begin{aligned}
    z_*= &\int \frac{dz}{1-v_z^2};  \\
    \Rightarrow z_*\sim &\frac{1}{2}\ln{\left(1-v_z^2\right)}\left(\frac{dv_z}{dz}\right)^{-1}
    \end{aligned}
\end{equation}

In terms of $z_*$, equation (14) reduces to a free Schr\"{o}dinger-like equation 
\begin{equation}
    -\frac{d^2 \psi (z)}{dz_*^2}=\omega ^2 \psi (z).
\end{equation}

Here, $z(z_*)$ is implicit. We notice that in restricting to (1+1) dimensions we have avoided the complications that would arise in terms of a potential from spherical harmonics.

The modes have the form
\begin{equation}
    \psi(z(z_*))= A e ^ {i \omega z_*}
\end{equation}

Expanding this in terms of (15) we get
\begin{equation}
     \psi(z(z_*)) \sim  \exp{\left[i \omega\left(\frac{\ln{\left(1-v_z^2\right)}}{dv_z/dz}\right)\right]}
\end{equation}

This logarithmic pile-up is a key feature of several derivations of the Hawking temperature including Hawking's initial derivation where it emerges as a consequence of integrating momentum modes, obtained from a (spacetime dependent) non-trivial dispersion relation \cite{hawking}.

Modes, including $v_z=1$ would correspond to the ones straddling through the horizon. However, we notice the integral $\int \frac{dz}{1-v_z^2}$ diverges at $v_z=1$. This, in fact, is a consequence of a singularity in our choice of coordinates. Further, although $z_*$ is well defined only beyond the horizon ($v_z < 1$), we will still use it crudely, to get the right form of the solutions. We circumvent the pole by following Feynman's $i \epsilon$ trick. That is, we focus on the following integral
$$
z_*=\int \lim_{\epsilon\to 0}\frac{dz}{1-v_z^2 + i \epsilon}
$$
Here, we make use of the Sokhotski–Plemelj theorem to convert this integral into the form
\begin{equation}
    z_*= \int dz \wp\left(\frac{1}{1-v_z^2}\right)-i \pi \delta\left(1-v_z^2\right)
\end{equation}

The delta function contribution will look like
$$
\frac{i \pi}{ {2v_z (dv_z/dz)} }
$$

This, when plugged into (17), will yield

\begin{equation}
    \psi(z(z_*)) \sim \exp{\left( \frac{\omega \pi}{dv_z/dz}  \right)} 
\end{equation}

This factor, which appears with the energy ($\omega$) of the mode, is in fact Boltzmann-like. (The contribution of the form $\exp(-2v_z)$, within the near horizon approximation, can be absorbed into the normalisation factor. It will thus show up in the relation between the normalisation factors in the straddling and non-straddling cases. See \cite{visserhawking} page 10.)

\subsection{Analogue Hawking temperature}

We take the above factor and define a quantity $T_H$ such that
\begin{equation}
    k_B T_H=\frac{K}{2 \pi} =\frac{1}{2 \pi} \frac{\partial v_z}{\partial z}.
\end{equation}

This quantity is the analogue of the Hawking temperature from black hole thermodynamics. Here $k_B$ is the Boltzmann constant. ($\hbar=1$)

We have, essentially, mimiced Visser's derivation of $T_H$ \cite{visserhawking}. However there is a crucial distinction. His paper involved treating the non-trivial wave vector $k$ as the central object for the derivation. On the other hand, we saw the same results emerge through coordinate transformations , particularly the form of the tortoise coordinate $z_*$ made the Hawking temperature manifest \cite{poisson}.

After having defined a hawking temperature, one would be tempted to define a corresponding analogue Bekenstein Entropy. However, as Visser had emphasised, in order to go from Hawking temperature (kinematics) to Bekenstein entropy (geometrodynamics), Einstein equations are required. Here, we see, that the acoustic analogue offers us a crucial distinction between two closely related ideas of black hole thermodynamics. However, there are extensions of the acoustic analogue, to neo-Newtonian frameworks for example, where an analogue definition of entropy can be defined \cite{ana2}.

\section{Analogue FLRW cosmology}
As Barcelo et.al. \cite{anaFLRW} pointed out, the appropriate acoustic analogue of an expanding FLRW metric, is obtained by keeping the background fluid at rest $(v_0=0)$ and instead varying the speed of the excitations ($c_s$) within them.

Starting with equation (5), again, two more conditions are imposed.

\begin{enumerate}
    \item The background fluid is considered to be at rest, $v_z=0$. The continuity equation gives that $\rho_0$ is a constant.
    \item Spatially homogenity is assumed. This implies that $c_s$ is independent of $z$.
\end{enumerate}
These lead to
\begin{equation}
   \mathrm{d} s^{2} =\frac{\rho_{0}}{c_s(t)}\left[- c_s^{2}\mathrm{~d} t^{2}+\mathrm{d} z^2 \right]. 
\end{equation}
Further, the metric can be rescaled by a constant factor $\frac{c_0}{\rho_0}$. Here $c_0$ is some convenient reference speed. The line element thus becomes
\begin{equation}
   \mathrm{d} s^{2} =\frac{c_{0}}{c_s(t)}\left[- c_s^{2}(t)\mathrm{~d} t^{2}+\mathrm{d} z^2 \right]=-c_0c_s(t)\mathrm{~d}t^2 + .\frac{c_{0}}{c_s(t)} \mathrm{d} z^2
\end{equation}
In order to convert this line element into the traditional FLRW form, a pseudo time is introduced. This pseudo time $\tau$ is related to the laboratory time $t$ such that 
\begin{equation}
    d \tau = dt \sqrt{c_s(t)/c_0}
\end{equation}

Equation (19) becomes
\begin{equation}
    \mathrm{d} s^{2} = -c_0^2 \mathrm{~d} \tau^{2} + \frac{c_0}{c_s(t)}\mathrm{d} z^2 = -c_0^2 \mathrm{~d} \tau^{2} + \frac{c_0^2}{\Bar{c}_s^2(\tau)}\mathrm{d} z^2
\end{equation}
Here, $\Bar{c}_s(\tau)$ is the speed of sound in terms of the pseudo time. This can be obtained via the chain rule
\begin{equation}
    \Bar{c}_s(\tau)= \frac{d z}{d \tau}= \frac{d t}{d \tau} \frac{d z}{d t} = \sqrt{c_0c_s(t)}
\end{equation}

Note, equation (26) is completely equivalent to the FLRW line element in (1+1) dimensions upon making the identification
$$
    c_0d \tau \longleftrightarrow c  dt_{F}  
$$ and
$$
\frac{c_0^2}{\Bar{c}_s^2(\tau)} \longleftrightarrow a^2(t_{F})
$$
Here $c$ is the speed of light and $t_F$ is the comoving FLRW time.
Note also that, however we restrict to the flat ($K=0$) case of the FLRW metric, one can look at the non-flat cases too. In fact in (1+1) dimensions the $K=\pm 1$ cases are related to the $K=0$ case by a trivial coordinate transformation of the form $dz^2 \xrightarrow{}dz^2/(1-Kz^2)$. This implies that in (1+1) dimensions, the choice of $K$ does not affect the evolution of the scale factor. \cite{mann}

In order to simplify the solutions, to a new time coordinate $\eta$ is used.
\begin{equation}
    d \eta = \frac{d \tau}{a(\tau)}= \frac{\Bar{c}_s(\tau)}{c_0} d \tau 
\end{equation}
This new time $\eta$ can be called the conformal pseudo time. In terms of $\eta$, equation (26) becomes
\begin{equation}
    \mathrm{d} s^{2} = a(\eta)[-c_0^2 d \eta^2 + dz^2] =\frac{c_0^2}{\Bar{c}_s^2(\tau(\eta))}[-c_0^2 d \eta^2 + dz^2]
\end{equation}

The general solutions for equation (2) corresponding to such space-times can be taken to be the modes
\begin{equation}
    \phi_k(z,\eta)= \frac{e^{i k z}}{\sqrt{(2 \pi)^3}} \psi_k(\eta)
\end{equation}
such that $\psi_k(\eta)$ will obey the equation 
\begin{equation}
    c_0^2\frac{d^2\psi_k(\eta)}{d \eta ^2} + k^2 \psi_k(\eta)=0
\end{equation}
Which gives
\begin{equation}
    \psi_k(\eta)\propto e^{\pm i \omega \eta}
\end{equation}
Here $\pm$ correspond to the incoming and outgoing modes and $\omega$ is fixed by the on-shell relation.

As mentioned previously, one cannot mimic the dynamics by which the scale factor $a(\tau)$ is obtained. However once the scale factor, or the corresponding $\Bar{c}_s(\tau)$ is obtained, it can plugged into (28) to relate the conformal time with the pseudo time.
For example, for a matter dominated universe $a(\tau)\propto \tau^\frac{2}{3}$. This would thus give us $\eta \sim \tau^\frac{1}{3}$ which could be plugged into (32).

Equation (32) tells us that the solutions are proportional to $e^{\pm i \omega \eta}$. This implies that the energy of the solutions do not change their signs. Thus no particles are created. This is to be expected, as it is known that particle creation does not take place for massless fields in conformally flat spacetimes. \cite{ford}. In the next subsection we will see how a particle production can take place within a spacetime which is not conformally flat.

\subsection{Parker-Toms Model}
Parker and Toms put forward a model, which allowed for particle production for massless and minimally coupled scalar fields in expanding spacetimes\cite{parker}. Here we sketch an acoustic correspondence with the Parker-Toms model.
Starting with equation (25) and introducing a new time coordinate $\tau_{pt}$ such that
\begin{equation}
    d \tau = \frac{c_0 ^3}{\Bar{c}_s ^3(\tau_{pt})}d \tau_{pt}
\end{equation}

Thus
\begin{equation}
    ds^2=  - \frac{c_0 ^6}{\Bar{c}_s ^6(\tau_{pt})} c_0 ^2d \tau^2_{pt} + \frac{c_0^2}{\Bar{c}_s^2(\tau_{pt})}dz^2
\end{equation}

The solutions again take the form
\begin{equation}
    \phi_k(z,\tau_{pt})= \frac{e^{i k z}}{\sqrt{(2 \pi)^3}} \psi_k(\tau_{pt})
\end{equation}
With $\psi_k(\tau_{pt})$ obeying
\begin{equation}
    \frac{d^2 \psi_k}{d \tau_{pt}^2} + k^2 \frac{c_0^4}{\Bar{c}^4_s(\tau_{pt})}c_0^2 \psi_k=0
\end{equation}

Note, unlike FLRW cosmology, where the scale factor is obtained by solving Freidmann's equations, in the Parker-Toms model, the scale factor is chosen by hand and such that the spacetime has staticaly bounded expansion. Following the prescription by Parker and Toms for the scale factor, the speed of sound in this new time co-ordinate is be chosen such that \cite{ford}

\begin{equation}
    \frac{c_0^4}{\Bar{c}_s^4(\tau_{pt})}= \frac{1}{2} \left[1 + \frac{c_0^4}{\Bar{c}_{s0}^4} +\left(1-\frac{c_0^4}{\Bar{c}_{s0}^4}\right) \tanh{(\alpha \tau_{pt})}\right]
\end{equation}
Here $\Bar{c}_{s0}$ is some initial speed of sound and $\alpha$ is some relevant energy scale. 

(31) becomes
\begin{equation}
    \frac{d^2 \psi_k}{d \tau_{pt}^2} + \frac{k^2}{2} \left[1 + \frac{c_0^4}{\Bar{c}_{s0}^4} +\left(1-\frac{c_0^4}{\Bar{c}_{s0}^4}\right) \tanh{(\alpha \tau_{pt})}\right] \psi_k=0
\end{equation}

This is completely equivalent to the results from QFT in curved spacetimes. For instance, in the high frequency limit $(k> \frac{kc_0^2}{\Bar{c}_s^2(\tau_{pt})} \gtrsim \alpha )$, the mean number of particles created in a mode $k$ can be calculated to be \cite{parkerbook}

\begin{equation}
    \left<N_k\right> \approx e^{-2 \pi k c_0^2/\Bar{c}_s^2 \alpha}
\end{equation}

For works exploring cosmological pair production within analogue cosmologies with emphasis on phenomenology, see \cite{Jain, lang, vissern}.


\section{Summary}
In this work, we highlighted the usefulness of the acoustic analogue of gravity to the study of QFTs in curved spacetimes. We did this by working in a model (1+1) dimensional acoustic spacetime and studying analogue versions of phenomena like Hawking temperature within black hole physics, and particle creation within cosmology. We also emphasised the limits of the acoustic analogue in describing the physics of semi-classical gravity, restricting only to the kinematical consequences of GR.

\section*{Acknowledgements}
I wish to thank Prof. Bijan Bagchi for the insightful discussions and encouragement in the early phases of this work. I also wish to thank Rahul Ghosh and Sauvik Sen for their valuable feedback.

\end{document}